\documentstyle[editedvolume,numreferences]{crckapb} 

\begin{opening}
\title{Basic Elements of Electrical Conduction}
\author{M.\ B\"uttiker and T. Christen}
\institute{D\'epartement de physique th\'eorique,\\
Universit\'e de Gen\`eve, \\
24, Quai E. Ansermet,
CH-1211 Gen\`eve 4, Switzerland}
\end{opening}
\runningtitle{Electrical Conduction}
\begin{document}
\section{Introduction}
\label{introduction}
Many phenomena of electrical conduction in small mesoscopic 
conductors have been successfully explained within the framework
of the scattering approach \cite{Review}. The main emphasis of
this work is an extension of this approach to time-dependent
phenomena. Of particular interest are the basic requirements which
a dynamic conductance of a mesoscopic conductor has to satisfy. There
is a gap in the way dynamic conduction 
is treated in a large body of the physics literature and in more
applied discussions based on simple circuit theory. For an
electric network we know that its time-dependent behavior is
crucially determined by its capacitive and inductive elements. 
In contrast to this, time-dependent phenomena are often treated 
by a response theory of non-interacting
carriers. Unfortunately, this approach to ac-conductance
which has been employed by some of the leading condensed matter
physicists \cite{MOTT} is prevalent.
We emphasize that in order to obtain a reasonable answer
it is not sufficient to consider non-interacting electrons,
a Fermi liquid or even a Luttinger liquid with short range
interactions. An analysis based on an electric circuit model gives an
answer which conserves the total charge and which
has the property that all frequency dependent currents at the input
and output nodes of such a circuit add up to zero.
To obtain results which are {\em charge and current conserving}, it is
necessary to consider the implication of the long range 
Coulomb interaction. This leads to a theory which can be considered to be
an extension of the work of Pines and Nozi\`eres \cite{PINES} and
Martin \cite{MARTIN} on bulk systems to mesoscopic conductors. 
Below we present a discussion \cite{MBCAP} which 
for simplicity focuses on the low-frequency transport. The benefit 
of this restriction is its considerable generality,
and that it can be used to analyze the self-consistency of the dc
scattering-approach (Sect. \ref{potential}),
the leading order nonlinearities of the 
current voltage characteristic (Sect. \ref{nonlinear}),
and the low-frequency dynamical conductance
(Sect. \ref{frequency}).\\ \indent
\begin{figure}
\vspace{6cm} 
\caption{Mesoscopic capacitor connected to contacts
with 
electrochemical potentials $\mu_{1}$ and $\mu_{2}$ and electrostatic potentials
$U_{1}$ and $U_{2}$. After Ref. 7.} 
\end{figure}
For a two-terminal conductor the low-frequency admittance is of the form 
\begin{eqnarray}
G(\omega) = G(0)-i\omega E_{\mu} + O(\omega^{2})\;\; .
\label{eqI1}
\end{eqnarray}
where $G(0)$ is the dc-conductance and $E_{\mu}$ is 
called the {\it emittance} \cite{MBCAP}.
The emittance describes the current response in the leading order
with respect to frequency and can be associated with the displacement
charge passing a contact. For conductors with poor transmission as,
e.g., a condenser or a metallic diffusive wire,
the emittance is positive: the current follows the voltage as is
characteristic of a capacitive response. On the other hand,
it turns out that the emittance is negative
in samples with good transmission.
The current leads the voltage as is characteristic of circuits 
with an inductance.

To illustrate the physics contained in
this response coefficient $E_{\mu}$, we present now a few results.
Their detailed derivation is presented in Sect.
\ref{examples}.\\ \indent
Consider first the case of a mesoscopic condenser \cite{WEISS,BTPC}
as shown in Fig. 1 with a
geometrical capacitance $C_{0}$. The measured electrochemical
capacitance $C_{\mu}$ is determined by the geometrical capacitance in 
series with quantum capacitances $e^{2} dN_{1}/dE$, $e^{2} dN_{2}/dE.$
Here $dN_{1}/dE$ and $dN_{2}/dE$ are the total densities of states of the
mesoscopic plates in the regions into which the electrical field penetrates.
Thus the electrochemical capacitance of a mesoscopic condenser is 
\cite{BTPC}
\begin{eqnarray}
C_{\mu}^{-1} = 
C^{-1}_{0} + (e^{2} dN_{1}/dE)^{-1}+ (e^{2}
dN_{2}/dE)^{-1}.
\label{cmu}
\end{eqnarray}
As it must be, in the macroscopic limit, where the
densities of states diverge, one finds $E_{\mu}=C_{0}$. On the other
hand, if the density of states at the surfaces of the plates are 
sufficiently small,
the geometrical capacitance
can be neglected. For the symmetric case this leads to the emittance
$E_{\mu}= (1/4)e^{2}(dN/dE)$, where $dN/dE = dN_{1}/dE+ dN_{2}/dE$ 
is the total density of states
of the two surfaces.\\ \indent
\begin{figure}
\vspace{6cm} 
\caption{Mesoscopic metallic conductor connecting two reservoirs.
The low-frequency response is capacitive and proportional to 1/6 of its
total density of states.}
\end{figure} 
Next consider a sample which permits transmission of carriers 
between the contacts. To be definite, first consider 
a metallic diffusive wire connecting two reservoirs (see Fig. 2).
Experiments have been carried out by Pieper and Price \cite{PIEP1}
and related theoretical work can be found in 
Refs. \cite{PIEP2,LIU,YSHAI,KRAMER}.  
We determine the local potential with the help of Thomas-Fermi screening
(a local charge-neutrality condition).
In the presence of an
applied dc-voltage, the ensemble averaged potential drops linearly.
The total density of states in the volume over which the electric
field is non-vanishing is denoted by $dN/dE$. We find 
that the emittance of such a metallic
diffusive wire is positive like a capacitance and is given by
\cite{MBUNP} 
\begin{eqnarray}
E_{\mu} = (1/6) e^{2} (dN/dE)\;\; .
\label{eqI2}
\end{eqnarray}
It is not the total density of states that counts!
Without screening and in the presence of a dc potential-drop, one-half
of the states would be filled. Screening, however, reduces the naively
expected result $(1/2) (dN/dE)$ by a factor $1/3$. Such a $1/3$-reduction
is familiar from the theory of shot noise of metallic diffusive conductors
were it is found that the actual shot noise is $1/3$ of the full
Poisson noise \cite{BEE13,NAGA,DEJON}.
The physical origin of the factor 1/3 in these two problems
is quite different but from a purely mathematical point of view the 
origin \cite{NAGA} of the factor 1/3 is the same.\\ \indent
Next consider a ballistic wire of length L. Now, for dc transport the
ensemble-averaged potential drops
as we enter the wire from the reservoir, it is
constant in the wire, and it drops again as we leave the wire. 
We find then a negative (inductive) emittance 
\begin{eqnarray}
E_{\mu} =  - (1/4) e^{2} (dN/dE)\;\;,
\label{eqI3}
\end{eqnarray}
where $dN/dE$ is the total density of states of the wire in the
volume where the potential is uniform.\\ \indent
As a further example, consider a resonant double barrier
\cite{PRICE,FU,MBHTP}. Suppose for simplicity that it is symmetric
and suppose that it is reasonable to determine the potential in the well 
from a charge neutrality condition. 
If we denote by $T$, $R=1-T$, and $dN/dE$
the transmission probability, the reflection probability, and the 
density of states in the well, respectively,
the emittance can be written in the form \cite{MBHTP,CHRIS}
\begin{eqnarray}
E_{\mu} = (1/4) (R-T) e^{2}(dN/dE)\;\;.
\label{eqI4}
\end{eqnarray}
Note that Eq. (\ref{eqI4}) interpolates between the emittance of
a symmetric condenser with large geometrical capacitance
and a ballistic wire. 
The emittance is negative (inductive) at resonance $T=1$ and crosses
zero at $T = R = 1/2$ where the Fermi level is a half-width of the
resonant level above (or below) the energy of the resonant state.
If the Fermi level is further than a half-width away from the resonant
level the emittance is positive (capacitive). The case of a double
barrier which is asymmetric has been considered in Ref. \cite{MBHTP}.
A result for single tunneling barriers which goes beyond Thomas-Fermi
screening will be discussed elsewhere \cite{CHRIS}.\\ \indent
\begin{figure}
\vspace{6cm} 
\caption{Hall conductors: (a) Corbino disk, (b)  Hall bar. After Ref. 22.} 
\end{figure} 
Finally, consider a two-dimensional electron gas
patterned into the shape of a Corbino disk (Fig. 3a)  or 
a Hall bar (Fig. 3b) with two metallic contacts.
In the range of magnetic fields over which the Hall conductance
is quantized the dc-resistances can be evaluated solely by considering
edge states \cite{MB88}. In a non-interacting theory the edge states intercept
the Fermi energy with a finite slope and have a local density of states
per unit length given by $dn(s)/dE =1/hv(s)$, where $s$ is a coordinate 
along the edge state and where $v(s)$ is the velocity of carriers at point
$s$. The integrated density of states along the edge channel $k$
is thus $dN_{k}/dE = \int ds (dn_{k}/dE)$. 
Suppose for simplicity that the magnetic field is such that we have 
only one pair of edge states. The Corbino disk acts then as an
insulator with vanishing dc conductance
$G(0) = 0$. The long-range Coulomb interaction between the inner and
the outer edge state can be described by a geometrical capacitance $C_{0}$
(which depends logarithmically on the width of the sample).
As one expects, we find that the Corbino-disk exhibits an
emittance which equals the electrochemical capacitance \cite{TCMB95} 
$E_{\mu} = C_{\mu}$, where $C_{\mu}$ is given by Eq. (\ref{cmu}).
A topologically different conductor of the same width and length but with
its contacts arranged at its ends is shown in Fig. 3b.
Now the two-terminal dc-conductance is quantized and given by
$G(0) = e^{2}/h$ for spin-split Landau levels. Obviously,
the low frequency response of this conductor is
not capacitive but dominated by transmission of carriers between
different reservoirs. In particular, we find 
the inductive-like emittance \cite{TCMB95} 
$E_{\mu}=-C_{\mu}$. Thus the ac-response of a
quantized Hall conductor is determined by the way the edge states
are connected to reservoirs. In the Corbino disk each edge state
returns to the reservoir from which it
emanates. In contrast, in the two-terminal Hall-bar, each edge state 
connects to a reservoir that is different from the one from which
it emanates. We do not discuss this example anymore in the sequel.
A forthcoming publication \cite{TCMB95} presents a
general formula for the emittance which is applicable to a wide
variety of samples with different edge-state topologies. 

\section{The Scattering Approach}
\label{transmission}
Consider a conductor connected to a number of contacts \cite{MB86}. 
The contacts are 
labeled with the greek indices $\alpha$. We assume that the distance 
between these contacts is so small that transmission from one contact 
to another one can be considered to be phase coherent. Thus we assume that
scattering inside the conductor is purely elastic. 
For small deviations of the electrochemical potentials away from their
equilibrium value the 
dc-current
$I_{\alpha}$ at probe $\alpha$ is given by \cite{MB86}
\begin{eqnarray}
I_{\alpha} = (e/h) \int dE (-\frac{df}{dE})
\left( (N_{\alpha} - R_{\alpha \alpha}) \mu_{\alpha}
- \sum_{\beta} T_{\alpha \beta} \mu_{\beta} \right) .
\label{eq01}
\end{eqnarray}
Here, $N_{\alpha}$ is the number of quantum channels 
with thresholds below the equilibrium electrochemical potential 
in contact $\alpha$, and $f$ is the Fermi function.
Carriers in the $N_{\alpha}$ 
incident channels have a total combined probability 
$R_{\alpha \alpha}$ for reflection back into contact $\alpha$.
Carriers incident in contact $\beta$ have a total probability
$T_{\alpha \beta}$ to traverse the sample into contact $\alpha .$ 
Equation (\ref{eq01}) is a quantum-mechanical Kirchhoff law. It states
the conservation of the current at an arbitrary intersection of 
a mesoscopic wire and that the currents are only a function of voltage
differences. In the present context
these features are a consequence of the unitarity 
of the scattering matrix ${\bf S}$ and its behavior under time reversal.
The fact that the scattering matrix is unitary and that
the microscopic equations are reversible implies that under a 
reversal of magnetic field the scattering matrix has the symmetry
${\bf S}^{T}(B) = {\bf S} (-B)$. The scattering matrix ${\bf S}$ for
the conductor can be arranged such that
it is composed of sub-matrices ${\bf s}_{\alpha \beta }$
which relate the incident current amplitudes in probe $\beta$ 
to the outgoing current amplitudes in probe $\alpha$.
In terms of these scattering matrices the probabilities introduced above 
are simply $T_{\alpha \beta} = Tr({\bf s}^{\dagger}_{\alpha \beta }
{\bf s}_{\alpha \beta })$ if $\alpha \ne \beta$ and 
$R_{\alpha \alpha} = Tr({\bf s}^{\dagger}_{\alpha \alpha}
{\bf s}_{\alpha \alpha})$. The trace is over the quantum channels in 
each probe. As a consequence of the unitary properties of the scattering
matrix, we have $N_{\alpha} = 
R_{\alpha \alpha} +\sum_{\beta} T_{\alpha \beta} $ and 
$N_{\alpha} = R_{\alpha \alpha} +\sum_{\beta} T_{\beta \alpha}$
which imply current conservation $\sum_{\alpha} I_{\alpha} = 0$.
Furthermore, the currents depend only on differences of the electrochemical
potentials.
 
Let us restate these properties in terms of conductance 
coefficients $G_{\alpha\beta} = I_{\alpha} /V_{\beta}.$ Taking into
account that in a reservoir the voltages and electrochemical potentials
move in synchronism, $\delta \mu_{\alpha} =eV_{\alpha}$, we have \cite{MB86}
\begin{eqnarray}
G_{\alpha\alpha}(0)  =\frac{e^{2}}{h}\int dE (- \frac{df}{dE})
(N_{\alpha} - R_{\alpha \alpha})
= \frac{e^{2}}{h} \int dE (- \frac{df}{dE})
\sum_{\beta} T_{\alpha \beta} \;\;.
\label{eq02}
\end{eqnarray}
for $ \alpha = \beta $ and      
\begin{eqnarray}
G_{\alpha\beta} (0) = - \frac{e^{2}}{h} \int dE (- \frac{df}{dE})
\: T_{\alpha \beta} \;\;.
\label{eq03}
\end{eqnarray}
Current conservation means for the conductance matrix  
\begin{eqnarray}
\sum_{\alpha} G_{\alpha\beta}(0)  =  0
\label{eq04}
\end{eqnarray}
for any $\beta$. On the other hand, the statement
that the currents can only depend on
voltage differences leads to 
\begin{eqnarray}
\sum_{\beta} G_{\alpha\beta}(0)  =  0\;\;.
\label{eq05}
\end{eqnarray}
We emphasize the simple properties (\ref{eq04}) and (\ref{eq05})
since later on we point out that they are also valid for 
$G_{\alpha\beta}(\omega).$ 

Before proceeding we briefly comment on closely related work.
In his work Landauer determined the voltage
drop across an obstacle with the help of a local charge neutrality
argument applied to the perfect sections of the conductor on either side 
of an obstacle \cite{LA57}. For an obstacle with transmission probability T,
this leads to a resistance proportional to $(1-T)/T$. This 
resistance has the obviously correct property that it vanishes
in the limit of perfect transmission. 
Already Engquist and Anderson pointed out that voltages
measured at contacts are determined not by a charge neutrality condition
but by adjusting the electrochemical potential of a voltage probe in such 
a manner that the voltmeter draws zero net current \cite{EA81}. 
Due to the special 
geometry considered, the restriction to a one channel conductor,
and the restriction to a phase-incoherent voltage measurement \cite{MB89}, 
the answer
found by Engquist and Anderson is the same as that found by Landauer. 
The key notion that the resistance across an obstacle should vanish
if the obstacle permits perfect transmission persisted long after the 
work of Engquist and Anderson. The task seemed, therefore, to be to find
resistance formulae \cite{AZ81} of the type $(1-T)/T$. 
In a work that largely centers
on this notion, Imry \cite{IM86} observed that if one considers in 
the discussion of  
Engquist and Anderson not the voltage drop across the obstacle but the
voltage drop between the current source and sink one obtains a resistance
that is proportional to $T^{-1}$, i. e. a conductance that is proportional to
$T$. 
Equation (\ref{eq01}), in contrast, was motivated by experiments by
Benoit et al. \cite{BE86}
in which the role of the current and voltage source was exchanged.
Such experiments imply that we need a formulation of electrical
transport in which apriori all contacts are treated equivalently and 
on an equal footing \cite{MB86}. In a specific arrangement of current 
and voltage sources
the voltages are a posteriori determined by a zero current condition. 
This leads to conductances which are electrochemical quantities
and which obey the basic requirements of a transport theory. These
basic requirements include a reciprocity symmetry \cite{MB86,BE86,VANH} (which
is a consequence of microreversibility and the irreversibility of thermodynamic
electron reservoirs) and include a fluctuation-dissipation theorem \cite{MB92}.
Below we discuss in more detail the electrochemical requirements 
which lead to Eq. (\ref{eq01}). 

It is sometimes implied in the
theoretical literature, that the derivation of conductance formulae is
just a question of applying formal linear response theory correctly to this 
problem. This is not borne out by the history of this field: Depending on
one's preconceived notions quite different results can be derived \cite{THOU}.
But it is certainly true, that the derivation of these results from
linear response has made them more acceptable. This is, however, neither a 
question of rigour nor depth but simply a consequence of the fact
that even simple 
results are often only accepted if they are embellished by a sufficiently 
complicated derivation. For very concise linear response discussions of 
Eq. (\ref{eq01}) and some still open related questions we refer the 
reader to Ref. \cite{SHEP,MBAPT,NOECK}.

\begin{figure}
\vspace{6cm} 
\caption{Electrochemical and electric potentials of a mesoscopic capacitor:
An increase of the electrochemical potential $\mu_{1}$ by $d\mu_{1}$ causes
the band bottom to shift by $edU({\bf r}) = u_{1}({\bf r})d\mu_{1}$ where
$u_{1}$ is the characteristic potential of contact $1$. 
After Ref. 5.}
\end{figure} 

\section{Potential Distribution in a Mesoscopic Conductor}
\label{potential}
The scattering approach as discussed above seems like
a simple black-box approach. If we know through which contact 
carriers leave the sample and know the current that is injected by a 
reservoir we can find the total currents. The issue which is not
trivial is the fact that the voltages at the contacts must be
well-defined in order for Eq. (\ref{eq02}) and Eq. (\ref{eq03})
to be the correct answer. In many articles on the subject one
finds a conductance attributed to a wire which is strictly
one-dimensional, or one finds pictures of 
mesoscopic conductors with leads that are narrower then the mesoscopic
sample itself and are called reservoirs. Such geometries, as we 
now show, do in fact not lead to the conductances given above. This
criticism is not novel: it has been aired in a number of papers by
Landauer \cite{LA87} and one of the authors \cite{MB92} - unfortunately without
much success. The following discussion is most closely related to work by
Levinson \cite{LE89} but in detail 
follows Ref. \cite{MBCAP}.\\ \indent
For Eqs. (\ref{eq02}) and (\ref{eq03}) to be valid, the electric potential
$U_{\alpha}$ in the reservoirs $\alpha $ must follow the electrochemical
potential $\mu_{\alpha}$ in this contact. To generate transport we must 
consider a non-equilibrium situation. The electrochemical potentials 
in the contacts $\mu_{\alpha}$ must be allowed to differ from the 
equilibrium chemical potential $\mu _{0}$. Suppose that the increment
of the electrochemical potential in contact $\alpha$ is 
$\delta \mu_{\alpha} = \mu_{\alpha} - \mu _{0}$. The electric potential 
changes from its equilibrium configuration $U_{eq}({\bf r})$ to
a new non-equilibrium configuration $U([\mu_{\alpha}],{\bf r})$.
Here the argument $[\mu_{\alpha}]$ indicates that the non-equilibrium
potential depends on the electrochemical potentials in the contacts.
Equations (\ref{eq02}) and (\ref{eq03}) presuppose 
that deep in contact  $\alpha$ the electrochemical and electric potential 
change in synchronism, i. e. $\delta \mu_{\alpha} = e\delta
U([\mu_{\alpha}],{\bf r})$. This is a consequence
of the charge neutrality deep in the reservoirs. 
The difference between the electrochemical potential and the 
electrostatic potential is the Fermi energy,
$E_{F\alpha} ({\bf r}) = \mu_{\alpha} -eU({\bf r})$.
This chemical potential $E_{F\alpha}$ determines the charge in the
neighborhood of ${\bf r}$. Deep in the reservoir the charge cannot
change, even if we bias the conductor. Consequently we have  
$0\equiv \delta E_{F\alpha} ({\bf r}) = \delta \mu_{\alpha} -e
\delta U([\mu_{\alpha}],{\bf r})$.
Note that we have taken here the chemical potential to be space 
depended. However, if the quantum coherence of the wave functions
is taken serious then also deep in the reservoir, i. e. in the wide wire,
a single energy $E_{F\alpha}$ is all that is needed to specify the
chemical potential.
The energy dispersions in a wide wire are of the type
$E_{\alpha n}(k) = \hbar^{2} k^{2}/2m + E_{\alpha n}^{0} + eU_{\alpha} $.
Here $\hbar^{2} k^{2}/2m$ is the longitudinal energy 
for motion along the wire, $E_{\alpha n}^{0}$ is the energy for transverse 
motion (the channel threshold)
and $eU_{\alpha}$ is the equilibrium potential (the bottom of the conduction
band) in contact $\alpha$. The energy dispersion depends in an explicit
manner only on a single spatially independent constant $eU_{\alpha}$. 
Instead of the spatially dependent relationship between the electrochemical,
chemical, and electrostatic potentials we have for coherent conductors
the relation $E_{F\alpha} = \mu_{\alpha} -eU_{\alpha} $. In either case 
the differential relationship is spatially independent in the lead.
In order to discuss the validity of Eqs. (\ref{eq01})-(\ref{eq03}),
we have thus to find the 
electrostatic potential and the conditions under which it 
changes synchronously with the electrochemical potential in the reservoirs.

\subsection{Characteristic Potentials}
\label{characteristic}
The electrostatic potential $U([\mu_{\alpha}], {\bf r})$ for mesoscopic 
conductors is a function of the electrochemical potentials of the contacts,
and a complicated function of position.
Small increases in the electrochemical potentials $\delta \mu_{\alpha}$
bring the conductor to a new state (see Fig. 4)
with an electrostatic potential
$U([\mu_{\alpha}+\delta \mu_{\alpha}],{\bf r})$.
The difference $\delta U$ between these two potentials
can be expanded in powers of the increment in the electrochemical potential.
To linear order we have
\begin{eqnarray}
e\: \delta U([\mu_{\alpha}], {\bf r}) =
\sum_{\alpha} u_{\alpha}({\bf r}) \: \delta \mu_{\alpha}\;\;.
\label{eq06}
\end{eqnarray}
Here, $u_{\alpha}({\bf r}) =$
$e \: \partial U([\mu_{\alpha}], {\bf r})/
\partial \mu_{\alpha}|_{\mu_{\alpha} = \mu_{0}}$, 
with $\alpha = 1, 2$ are the {\it characteristic
potentials} \cite{MBCAP} which determine the electrostatic potential
inside the sample in response to a variation of an electrochemical
potential at a contact.\\ \indent
Suppose for a moment that we increase all electrochemical potentials 
simultaneously and by the same amount, $\delta \mu _{\alpha }\equiv
\delta \mu$. Both before and after the change
the conductor is at equilibrium, hence the physical
properties of the conductor remain unchanged.

Consequently, the shift of the electrochemical potentials must be
accompanied by a shift $e\delta U \equiv  \delta \mu$ of the electrical
potential. Imposing this condition on Eq. (\ref{eq06}) implies that
the sum of all characteristic potentials is equal to one at every
space point \cite{MBCAP},
\begin{eqnarray}
\sum_{\alpha} u_{\alpha}({\bf r}) \equiv 1\;\;.
\label{eq07}
\end{eqnarray}
Equation (\ref{eq07})
is a consequence of the long-range Coulomb interaction.
It is the most important result of this work:
the conservation of 
charge under the application of a dc or ac bias and the conservation
of current are, as we will show, a consequence of Eq. (\ref{eq07}). 
\\ \indent
Let us now return to our original problem and consider what happens
if we increase just one electrochemical potential, say in reservoir
$\alpha $, by $\delta \mu _{\alpha }$. Obviously, the condition that 
the electrochemical potential and the electrostatic potential
move in synchronism deep inside reservoir $\alpha$ implies 
that the characteristic function $u_{\alpha}({\bf r})=1$
for ${\bf r}$ deep inside reservoir $\alpha$.
Together with Eq. (\ref{eq06}), this implies that Eqs.
(\ref{eq01})-(\ref{eq03}) are valid if and only if 
the characteristic potentials have the property that
$u_{\alpha}({\bf r}) = 1$ for ${\bf r}$ deep in contact 
$\alpha$ and $u_{\alpha}({\bf r}) = 0$ for ${\bf r}$ deep in any
other contact.\\ \indent
The electrostatic potentials are determined by the charge
distribution in the sample. As we increase the chemical potential
of contact $\alpha$ {\em keeping all electrostatic potentials fixed},
the additional charge 
\begin{eqnarray}
\delta n({\bf r}) =
(dn({\bf r}, \alpha)/dE) \: \delta \mu_{\alpha} 
\label{eq08}
\end{eqnarray}
enters the conductor.
Here, $dn({\bf r}, \alpha)/dE$ is the {\it injectivity} of contact
$\alpha$ into point ${\bf r}$ of the sample.
With the help of the scattering states $\Psi_{\alpha n}({\bf r})$
which have unit amplitude in the incident channel $n$
in lead $\alpha$, the injectivity can be expressed as
\begin{eqnarray}
dn({\bf r}, \alpha)/dE = \sum_{n}  (hv_{\alpha n})^{-1} 
|\Psi_{\alpha n}({\bf r})|^{2}
\label{eq09}
\end{eqnarray}
where $v_{\alpha n}$ is the velocity of carriers at the 
Fermi energy in channel $n$ in contact $\alpha$. Equation (\ref{eq08})
gives of course not the true density variation. The injected charges
induce a change in the electrostatic potential which
in turn implies an induced contribution to the density, $\delta
n_{ind}$,  which has to
be determined self-consistently. The total charge density is 
\begin{eqnarray}
\delta n({\bf r}) =
(dn({\bf r}, \alpha)/dE)\: \delta \mu_{\alpha} + \delta n_{ind}({\bf r})\;\;.
\label{eq10}
\end{eqnarray}
The induced charge density is connected to the
electrostatic potential via the response function $\Pi({\bf r},
{\bf r}^{\prime})$ (Lindhard function):  
\begin{eqnarray}
\delta n_{ind} ({\bf r}) = - \int d^{3}r^{\prime} \:
\Pi({\bf r}, {\bf r}^{\prime})\: e \delta U({\bf r}^{\prime})\;\;.
\label{eq11}
\end{eqnarray}
The response function can be expressed in terms of the scattering states
(Green's function of the Schr\"odinger equation). Note that the
Lindhard function describes the variation of the
charge density not only of the mobile electrons which can be
reached from the contacts but also of the localized states.
For the purpose of our discussion we simply assume that this response 
function has been calculated and is known. There is one property
of the Lindhard function which is needed later on and which is a simple 
consequence of the invariance of the electrical system under a global
potential shift. A simultaneous change 
in all electrochemical potentials injects a charge
\begin{eqnarray}
 \delta n({\bf r}) = \sum_{\alpha} 
(dn({\bf r}, \alpha)/dE) \delta \mu + \delta n_{ind}({\bf
r})\equiv 0 \;\;.
\label{eq12}
\end{eqnarray}
Taking into account Eq. (\ref{eq11}) and $e\delta U = \delta \mu$,
one concludes \cite{MBCAP}
\begin{eqnarray} 
dn({\bf r})/dE  =  \int d^{3}r^{\prime} \;\;
\Pi({\bf r}, {\bf r}^{\prime}) \;\;.
\label{eq13}
\end{eqnarray}
The quantity on the left hand side side of 
Eq. (\ref{eq13}),
\begin{equation}
dn({\bf r})/dE= \sum_{\alpha} 
(dn({\bf r}, \alpha)/dE)\;\;,
\label{ldos}
\end{equation}
is the local density of states
which equals the sum of all injectivities. Equation (\ref{eq13})
connects a chemical response quantity, the local density of states,
to the Lindhard function $\Pi$. It is in this
regard similar to the Einstein relation between a diffusion constant
and a conductivity. \\ \indent
Now we come back to the case where the voltage is changed only in contact
$\alpha $. By inserting Eq. (\ref{eq10})
into Poisson's equation and taking $e \delta U({\bf r})
= u_{\alpha}({\bf r}) \delta \mu_{\alpha}$ into account, we find
that the characteristic potential $u_{\alpha}({\bf r})$ is
the solution of a field equation with a non-local screening kernel
and a source term given by the {\it injectivity}
of contact $\alpha$,
\begin{eqnarray}
-\Delta u_{\alpha} ({\bf r}) + 4 \pi e^{2}
\int d^{3}r^{\prime}
\Pi ({\bf r}, {\bf r}^{\prime})
u_{\alpha} ({\bf r}^{\prime}) =
4 \pi e^{2} (dn({\bf r},\alpha )/dE) \;\;.
\label{eq14}
\end{eqnarray}
We define the Green's function $g({\bf r}, {\bf r}_{0})$
as the solution of Eq. (\ref{eq14}) 
with the source term $e\: dn({\bf r},\alpha)/dE$ replaced
by a localized test charge $ e\delta({\bf r}-{\bf r}_{0})$
at point ${\bf r}_{0}$. The characteristic potential $u_{\alpha}({\bf
r}) $ can then be written in the form 
\begin{eqnarray}
u_{\alpha} ({\bf r}) =
\int d^{3}r^{\prime} \: g ({\bf r}, {\bf r}^{\prime})\:
(dn({\bf r}^{\prime},\alpha )/dE) \;\;.
\label{eq15}
\end{eqnarray}
Using Eqs. (\ref{eq07}) and (\ref{ldos}), a summation over $\alpha $
implies for Green's function the property \cite{MBCAP}
\begin{eqnarray}
\int d^{3}r^{\prime} g ({\bf r}, {\bf r}^{\prime})
\sum_{\alpha} (dn({\bf r}^{\prime},\alpha )/dE) = 
\int d^{3}r^{\prime} g ({\bf r}, {\bf r}^{\prime})\:
(dn({\bf r}^{\prime})/dE ) \equiv 1 \;\;.
\label{eq16}
\end{eqnarray}
The same relationship follows from the condition that the sum of all induced
charge densities plus the test charge is zero.\\ \indent 
Now we find the condition for the electrical
self-consistency of Eqs (\ref{eq01}). According to Eq. (\ref{eq16})
the characteristic potential is equal to unity if the Green's function is
convoluted with the local density of states. Therefore, 
we must have that the injectivity  
$dn({\bf r},\alpha )/dE$ deep in contact $\alpha$ is
equal to the local density of
states $dn({\bf r})/dE$. This requires that nearly all (in a
thermodynamic sense) electrons
approaching the contact $\alpha$ be reflected into the reservoir.
If the conductor and the reservoir consist of the same material then
the reservoir must be wide compared to the mesoscopic conductor.
In semiconductor samples with metallic contacts, on the other hand,
the contact might be actually narrow compared to the dimensions of
the semiconductor since the density of states at the Fermi energy
of the metal is much larger then that of the semiconductor. This
is the case, for instance, in Ga/As-samples used to measure the
quantized Hall effect \cite{MB88}. Our emphasis that self-consistency requires
geometries which are of a wide-narrow-wide 
geometry deserves further discussion:
the notion that a portion of length $L$ of a purely one-dimensional
conductor has a conductance $G = (e^{2}/h) T$ (per spin)
seems to be widely accepted.
In contrast, from the point of view taken here a strictly one-dimensional
conductor cannot be characterized by a conductance.

\subsection{Charge Conservation}
\label{conservation}
Let us now demonstrate that if the conditions of self-consistency are
met, then the application of an external bias to the conductor preserves
the total charge in the system. To be more precise, imagine a volume
$\Omega$ which encloses the entire conductor including
a portion of the reservoirs which is so large that at the place
were the surface of $\Omega$ intersects the reservoir all the
characteristic potentials are either zero or unity. We demonstrate
charge conservation for the case that we 
increase $\mu_{1}$ by a small value $\delta \mu_{1}$ above the equilibrium 
chemical potential. The change in density is given by Eq. (\ref{eq10}).
Expressing the induced charge with the help of the Lindhard function
$\Pi({\bf r},{\bf r^{\prime}})$ and using
the Green's function to relate the characteristic potentials to the 
injected density gives
\begin{eqnarray} 
\delta n({\bf r}) = \left(
\frac{dn({\bf r}, 1)}{dE} - 
\int d^{3}r^{\prime} d^{3}r^{\prime \prime}
\Pi({\bf r}, {\bf r}^{\prime}) g ({\bf r}^{\prime}, {\bf r}^{\prime \prime})
\frac{dn({\bf r}^{\prime \prime},1 )}{dE} \right) \delta \mu_{1}
\label{eq17}
\end{eqnarray}
The total variation in charge is
$\int_{\Omega} d^{3}r \: \delta n({\bf r}) =$ 
\begin{eqnarray} 
\int_{\Omega} d^{3}r^{\prime}
\left( \frac{dn({\bf r^{\prime}}, 1)}{dE} - 
\int_{\Omega} d^{3}r^{\prime \prime}
\frac{dn({\bf r}^{\prime})}{dE}
g ({\bf r}^{\prime}, {\bf r}^{\prime \prime})
\frac{dn({\bf r}^{\prime \prime},1)}{dE} \right) \delta \mu_{1}
\label{eq18}
\end{eqnarray}
where we used Eq. (\ref{eq13}) and the symmetry property
$\Pi({\bf r},{\bf r^{\prime}}) = \Pi({\bf r^{\prime}},{\bf r})$ of the
Lindhard function. Equation (\ref{eq16}) then
implies that the total variation in charge inside the
volume $\Omega $ vanishes.
This can be understood in the following way.
According to the law of Gauss the charge included in a volume $\Omega$ is 
$\int_{ \partial \Omega} {\bf E} d{\bf S} = 4 \pi Q$. The charge in
$\Omega$ is conserved if the electric flux through the surface of
$\Omega$ vanishes. This means that for this conductor all electric
field-lines 
which are generated when we bias the sample have their sources and 
sinks within the volume $\Omega$. Application of a bias voltage to an 
electrical conductor results in a redistribution of the charge 
within our sample but not in an overall change of the charge.
If the conductor is poor, i. e. nearly an insulator, the reservoirs act
like plates of capacitors. In this case long-range fields exist which 
run from one reservoir to the other and from a reservoir to a portion of
the conductor. But if we chose the volume $\Omega$ to be large enough
then all field lines stay within this volume.\\ \indent 
In all these considerations we have implicitly assumed that our conductor
and the reservoirs is all that counts. Such a situation might be realized
for the metallic mesoscopic structures fabricated on insulating substrates.
But this picture is certainly not complete if we deal with modern 
mesoscopic semiconductor structures which are often defined with the help
of a number of nearby gates. In such a case we must take a broader view 
and include inside our volume $\Omega$ not only the conductor of interest
but also all of the nearby gates \cite{MBCAP}. 
>From an electrostatic point of view,
we deal then not only with
the mesoscopic object of interest but we have to take into account the
nearby electrical bodies used to define this object. In such a case the 
overall charge is still conserved, even though the total charge 
on the mesoscopic conductor of interest varies with the applied bias. 
The theory presented here can easily be extended to this case \cite{MBCAP}. 

\section{Nonlinear I-V Characteristic}
\label{nonlinear}                                                                                
As an application of the discussion given above, let us consider
the nonlinear I-V characteristic. The discussion presented 
here can also be carried out in terms of an external and internal
response and provides a nice illustration of these concepts \cite{MBCAP}.  
Nonlinearities in metallic mesoscopic samples have been analyzed by              
Al'tshuler and Khmelnitskii \cite{KHME} using diagrammatic techniques             
without a self-consistent potential.                                            
For transmission through a tunnel contact                                       
the effect of a potential which changes with increasing                         
applied bias has been investigated by Frenkel \cite{FREN}.                                                             
Landauer has pointed to the necessity of a self-consistent                      
treatment of the internal potential \cite{LAND}.                                 
In this section we derive the current-voltage characteristic taking                       
into account nonlinearities which are a consequence                            
of the increase of the external                                                 
electrochemical potential differences as well                                  
as the changing internal (self-consistent) potential distribution.              
We focus on the leading nonlinear correction of                                
the low-voltage ohmic behavior of the sample.                                   
The I-V characteristic of a                                                     
mesoscopic sample is in general rectifying, i. e. $I(V) \ne - I(-V)$.                            
Furthermore, since rectification also depends on the internal                   
potential and since the internal potential in                                   
conductor $k$ depends on the charge distribution of other                       
nearby conductors, the rectification properties of a small                      
sample are dependent on its entire electric environment. 
Nevertheless, we consider for simplicity a conductor which is 
in electric isolation. Reference \cite{MBCAP} presents a more general result
being valid if there are additional nearby conductors like
gates or capacitors.\\ \indent                       
To proceed we view the scattering matrices                    
as a functional of the potential distribution                                   
${\bf s}_{\alpha\beta}(E,U([\mu_{\alpha}], {\bf r}))$ and expand $U$
away from the equilibrium potential-distribution. The scattering
matrix in the neighborhood of the equilibrium reference-state
(index $0$) is ${\bf s}_{\alpha\beta}(E, eU({\bf r}))                                                                    
={\bf s}^{0}_{\alpha\beta}(E) + \int d^{3}r^{\prime}                                                           
(\delta {\bf s}^{0}_{\alpha \beta}/                                                  
e \delta U({\bf r^{\prime}})) e\delta U({\bf r^{\prime}}).$                                                         
Here, $e \delta U({\bf r^{\prime}})$ can be expressed in terms                              
of the characteristic potentials and the                               
electrochemical potentials of the reservoirs.                                                   
The total current at probe $\alpha $ can be found by the 
same considerations that lead to Eq. (\ref{eq01}). However, we stop 
short of linearizing the resulting expression in the electrochemical 
potentials. The neglect of any inelastic scattering 
in the presence of large applied voltages is of course a limitation,
but since we only focus on the quadratic term in the voltages
this limitation might be not be so serious. The current is                                       
\cite{MB92}
\begin{eqnarray}                                                                
I_{\alpha} =                                                                 
(e/h)                                                                           
\sum_{\beta}\int dE f_{\beta}                                        
Tr \left[{\bf 1}_{\alpha}                                                   
\delta_{\alpha\beta}                                                            
- {\bf s}^{\dagger}_{\alpha\beta}(E, U({\bf r}))                               
{\bf s}_{\alpha\beta}(E, U({\bf r}))                                           
\right] \;\;,                                                                         
\label{eq30}                                                                    
\end{eqnarray}
where $f_{\beta }$ is the Fermi function belonging to reservoir
$\beta $. The sum over all currents at all terminals is still zero
due to the unitarity of the scattering matrix, hence the current
is conserved. 
In order that the current depends on voltage differences only,
it is necessary to treat the potentials self-consistently.
We expand Eq. (\ref{eq30}) in powers of                          
the electrochemical potential deviations $\delta \mu_{\alpha}=
eV_{\alpha}$,                   
\begin{eqnarray}                                                                
I_{\alpha} = \sum_\beta g_{\alpha\beta} V_{\beta}                      
+ (1/2) \sum_{\beta\gamma} g_{\alpha\beta\gamma} V _{\beta}         
V_{\gamma} \;\;.                                                            
\label{eq31}                                                                    
\end{eqnarray}                                                                  
The terms linear in the electrochemical potentials                             
are determined by the dc-conductances                                           
$g_{\alpha\beta} = (e^{2}/h) \int (-df/dE) Tr \left[{\bf 1}_{\alpha\beta}                                                  
- {\bf s}^{\dagger}_{\alpha\beta}(E)                                           
{\bf s}_{\alpha\beta}(E)                                                      
\right]$                                                                        
which are a functional of the equilibrium reference-potential only.             
The leading order nonlinear terms are given by transport                       
coefficients which are composed of an external and                     
an internal response:
\begin{equation}
g_{\alpha\beta\gamma} =                                                     
g^{e}_{\alpha\beta\gamma} +                                                  
g^{i}_{\alpha\beta\gamma} \;\; .
\label{gige}                                                 
\end{equation}
The external response arises from the expansion of the Fermi                    
functions in powers of the electrochemical potentials and is 
given by \cite{MBCAP}
\begin{eqnarray}                                                                
g^{e}_{\alpha\beta\gamma} =                                                  
- \frac{e^{3}}{h}                                                                                                                                            
\delta_{\beta\gamma}                                                            
\int dE (-\frac{df_{0}}{dE})                                                   
Tr \left[                                                                       
{\bf s}^{\dagger}_{\alpha\beta}                                             
\frac{d{\bf s}_{\alpha\beta}}{dE}                                           
+\frac{d{\bf s}^{\dagger}_{\alpha\beta}}{dE}                                
{\bf s}_{\alpha\beta}                                                       
\right]                                                                         
\label{eq32}                                                                    
\end{eqnarray}
where $f_{0}$ is the equilibrium Fermi function.
Let us examine the external response for the case that we have 
a two-terminal conductor.
>From the unitarity of the scattering matrix we find that all the
non-vanishing 
second-order conductance coefficients are equal in magnitude,
\begin{eqnarray}                                                                
g^{e}_{111} = - g^{e}_{122} =  - g^{e}_{211} = g^{e}_{222} \;\; .
\label{eq33}                                                                    
\end{eqnarray}
Furthermore, these coefficients can also be expressed just as the energy
derivative of the transmission probability,
\begin{eqnarray}                                                                
g^{e}_{111} = \frac{e^{3}}{h} \int dE (-\frac{df_{0}}{dE})
\frac{dT}{dE}        
\label{eq34}                                                                    
\end{eqnarray}
Thus up to the second order the contribution of the external
response to the current is
\begin{eqnarray}
I_{1} = - I_{2} = g (V_{1} - V_{2}) + (1/2) g^{e}_{111}
(V^{2}_{1} - V^{2}_{2})\;\;.          
\label{eq35}                                                                    
\end{eqnarray}
Despite the fact that currents are conserved this is an unphysical result.
The quadratic term depends not only on the voltage difference but 
on the individual voltages. Equation (\ref{eq35}) would predict that 
we should observe a different current depending on whether we rise
the voltage of the left contact by $\delta \mu_{1} = eV_{1}$
or whether we decrease
the voltage on the right contact by $ \delta \mu _{2}= - eV_{1}$
as compared
to the equilibrium value of the electrochemical potential.
This is a simple example which demonstrates
why the calculation of a nonlinear
current voltage characteristic without the self-consistent
adjustment of the electrostatic potential makes no sense.\\ \indent 
Let us now consider the internal response.
The internal response is a consequence of the change in                         
the potential distribution and is given by \cite{MBCAP}                                      
\begin{eqnarray}                                                                
g^{i}_{\alpha\beta\gamma} =                                                  
-\frac{e^{3}}{h}                                                                                                                                  
\int dE \: (-\frac{df_{0}}{dE})                                                   
\int d^{3}r                                                                     
Tr \left[                                                                       
{\bf s}^{\dagger}_{\alpha\beta}                                             
\frac{\delta                                                                    
{\bf s}_{\alpha\beta}}                                                      
{\delta eU({\bf r})}                                                            
+ \frac{\delta                                                                  
{\bf s}^{\dagger}_{\alpha\beta}}                                            
{\delta eU({\bf r})}                                                            
{\bf s}_{\alpha\beta}                                                       
\right]                                                                         
u_{\gamma} ({\bf r})                                                           
\nonumber \\                                                                      
- \frac{e^{3}}{h}                                                                                                                                 
\int dE \: (-\frac{df_{0}}{dE})                                                   
\int d^{3}r                                                                     
Tr \left[                                                                       
{\bf s}^{\dagger}_{\alpha\gamma}                                            
\frac{\delta                                                                    
{\bf s}_{\alpha\gamma}}                                                     
{\delta eU({\bf r})}                                                            
+ \frac{\delta                                                                  
{\bf s}^{\dagger}_{\alpha\gamma}}                                           
{\delta eU({\bf r})}                                                            
{\bf s}_{\alpha\gamma}                                                      
\right]                                                                         
u_{\beta}({\bf r})\;.                                                            
\label{eq36}                                                                    
\end{eqnarray} 
Note that the internal response contributes only to quadratic
order in the voltage.                                                                 
The linear conductance is a purely {\it external} response. If we now add
external and internal response, take into account Eq. (\ref{eq07})
and that the integral over $\Omega$ of an internal 
response term with $u = 1$ is equal to minus the external response
with the functional derivative replaced by an energy derivative,
we find \cite{MBCAP}
\begin{eqnarray}  
\sum_{\alpha} g_{\alpha\beta\gamma} =                                   
\sum_{\beta} g_{\alpha\beta\gamma} =                                
\sum_{\gamma} g_{\alpha\beta\gamma} = 0\;\; . 
\label{eq37}                                                                    
\end{eqnarray}
For a two-terminal conductor the second-order
conductance coefficients obey $ g_{111} = $ $- g_{112} =$
$  - g_{121} = $ $g_{122} = $
$- g_{211} = $ $g_{212} =$ $  - g_{221} = $ $- g_{222}$ 
Consequently, the currents are
\begin{eqnarray}
I_{1} = - I_{2} = g (V_{1} - V_{2}) + (1/2) g_{111} (V_{1} - V_{2})^{2}
+...  \;\;.        
\label{eq39}                                                                    
\end{eqnarray}
Now, the current depends only on the voltage difference as it must be.
In contrast to the external response which could simply
be expressed in terms of energy derivatives of the transmission
probability, the total response depends on the charge distribution
inside the conductor.\\ \indent
As a simple example, we consider an asymmetric resonant double 
barrier. The long lived state has a decay width $\Gamma_{1}$ to the 
left and $\Gamma_{2}$ to the right. 
For simplicity, assume that the potential in the well is determined by 
a local charge-neutrality argument (see Sec. 6). The characteristic potentials
in the well are $u_{1} = \Gamma_{1}/\Gamma$ and $u_{2} = \Gamma_{2}/\Gamma,$
where $\Gamma = \Gamma_{1} + \Gamma_{2}$ is the decay width.
>From Eqs. (\ref{eq32}) and (\ref{eq36}) we find for the second
order conductance coefficient
$g_{111} = ({e^{3}}/{h}) (dT/dE) (1-2u_{1})$ and hence \cite{MBUNP,CHRIS}
\begin{equation}
g_{111}= ({e^{3}}/{h}) (dT/dE) (\Gamma_{2} - \Gamma_{1})/\Gamma \;\;.
\label{g111}
\end{equation}
In summary, we emphasize that the nonlinearity cannot be discussed without
a concern for the way the potential drops in the interior of
the conductor.

\section{Frequency Dependent Conductance}
\label{frequency}
\subsection{External Response}
\label{external}

We are interested in the dynamical response of the conductor.
A time-dependent
voltage $ \propto \exp(-i\omega t)$
can be applied across two terminals, between a terminal and a
nearby gate, or between two nearby gates. We want to know the currents
which appear as a consequence of these oscillating voltages at the contacts
of the conductor or at the contacts to the nearby gates.
We are seeking the admittance matrix 
\begin{eqnarray}
G_{\alpha\beta}(\omega) =  I_{\alpha}(\omega)/ V_{\beta}(\omega)\;\;. 
\label{eq40}                                                                    
\end{eqnarray}
Again we consider for simplicity the case of a two-terminal
conductor in electrical isolation. The case where there are a number
of nearby gates (capacitors) or other conductors has been the
subject of a number of discussions \cite{MBCAP,MBROM}. 
Thus the indices  
$\alpha$, $\beta$ take the values $1$ and $2$ for the left and right
contact, respectively. If there are no other nearby electrical conductors
then all electric field lines which emanate from the conductor also
return to the conductor or to the reservoirs. Even in the presence
of time-dependent voltages applied to this conductor the reservoirs
remain locally charge neutral, i. e. the electric field
lines emanate from a reservoir only in the region where the
transition to the conductor occurs. If the conductor is very short 
then the reservoirs act like the plates of a capacitor and, 
due to long range Coulomb intercation, field lines connect the surfaces
of the two reservoirs facing each other. As in the dc-case 
there exists, therefore, a volume $\Omega$ which is so 
large that there is no electric flux through its surface. Consequently,
if we include all components of the system within $\Omega$ then the 
total charge within this volume is zero, i. e. all currents
at the terminals must add up to zero. Furthermore, 
since a potential which is uniform over the entire volume $\Omega$ 
is of no physical consequence the resulting currents must depend on
the potential differences only. Therefore, Eqs. (\ref{eq04}) and
(\ref{eq05}) hold also for the dynamic conductance: the rows
and columns of the dynamic conductance matrix $G_{\alpha
\beta} (\omega)$ must add up to zero.
We call such a discussion of the ac-transport a {\it charge} and 
{\it current conserving} theory. Below we illustrate the features of such a 
theory
for the case of low frequencies only. But the extension to a larger
range of frequencies and to nonlinearities must follow
the very same line of thought.\\ \indent

\subsection{Decomposition of the density of states}
\label{decomposition}
The total density of states in the conductor 
inside the volume $\Omega$ is a sum of {\it four}
contributions \cite{MBAPT},
\begin{eqnarray}
dN/{dE} = \sum_{\alpha\beta} (dN_{\alpha\beta}/{dE})\;\;,
\label{eq42}
\end{eqnarray}
where
\begin{eqnarray}
\frac{dN_{\alpha\beta}} {dE} =
\frac{1}{4\pi i}\,
{\rm Tr} \left[ {\bf s}^{\dagger}_{\alpha\beta} 
\frac{d{\bf s}_{\alpha\beta} }{dE}
-\frac{d{\bf s}^{\dagger}_{\alpha\beta} }{dE}
{\bf s}_{\alpha\beta}\right] 
\label{eq43}
\end{eqnarray} 
are the {\it partial densities of states}. 
Fig. 5 gives a schematic representation of the partial density
of states. The partial density of states $dN_{11}/dE$ consists of 
carriers that originate in contact $1$ and return to contact $1$.
The partial density of states $dN_{21}/dE$ consists of carriers
that originate in contact $1$ and are transmitted to contact $2$.
It turns out that 
the external response is determined exactly
by these four partial densities of states.
The partial densities of states represent a decomposition
of the total density of states both with respect to the 
origin of the carriers (injecting contact, right index)
and the final destination of the carriers (emitting contact, left
index).\\ \indent
\begin{figure}
\vspace{6cm}
\caption{Decomposition of the total density of states of a two 
terminal conductor. After Ref. 13.}
\end{figure}
At {\em constant electrostatic potential} the total charge injected into 
the conductor under a simultaneous and equal 
increase of the chemical potentials 
at its contacts $\delta E_{F1} = \delta E_{F2} = e \delta \mu$ is 
$\delta Q ^{e} = e \sum_{\alpha\beta} 
(dN_{\alpha\beta}/dE) \delta \mu$. To find the current at contact
$\alpha$ we need to know which portion of this charge enters or leaves the
conductor through contact $\alpha $, i. e. how the total charge
is partitioned on the two contacts. The answer to this question was
found by one of the authors, Pretre and Thomas
\cite{MBAPT} using a linear response calculation. The following
simple argument leads to the same result.
The scattering matrix ${\bf s}_{\alpha\beta}$ determines
the current
amplitudes of the outgoing waves in contact $\alpha$ as a
function of the current amplitudes
of the incident waves in contact  $\beta$. The charge which is
injected by an increase of the Fermi energy at contact $1$ is
$(e dN_{11}/{dE}+
e dN_{21}/{dE})
e V_{1}(\omega)$. 
Only the additional charge
$\delta Q_{1}^{e}(\omega) = e (dN_{11}/{dE}) e V_{1}(\omega)$ 
leads to a current at contact 1,
whereas $\delta Q_{2}^{e}(\omega) = e (dN_{21}/{dE})
e  V_{1}(\omega)$ is determined by carriers which leave
the conductor through contact $2$. Therefore, a variation of the 
Fermi levels of the contacts $\beta $ causes the current
\begin{eqnarray}
\delta I_{\alpha}^{e}(\omega) = -i \omega e^{2} \sum _{\beta }
(dN_{\alpha\beta}/{dE})
V_{\beta}(\omega)
\label{eq44}
\end{eqnarray}
at contact $\alpha $. Since direct transmission between contact
$1$ and $2$ is possible, an oscillating voltage causes in addition
at these contacts a current determined by the dc-conductance.
Thus the leading {\em external} low-frequency current response
to an oscillating chemical potential
$eV_{\alpha}(\omega)$ is given by \cite{MBAPT}
\begin{eqnarray}
G^{e}_{\alpha\beta}(\omega) = G_{\alpha\beta}(0)
-i \omega e^{2} (dN_{\alpha\beta}/{dE})\;\;.
\label{eq45}
\end{eqnarray}
This external response is not current conserving.
Since the dc-conduc\-tan\-ces satisfy 
$\sum_{\alpha} G_{\alpha\beta}(0)=
\sum_{\beta} G_{\alpha\beta}(0)=0$,
one finds that to leading order in frequency 
$\sum_{\alpha} G_{\alpha\beta}^{e}(\omega)$
is proportional to the total charge injected from contact $\beta$
into the conductor. The injected charges create an internal,
time-dependent electric potential
$\delta U({\bf r}, t)$ which in turn causes additional currents.
In the following subsection, we investigate the response to such
an internal electrostatic potential. It turns out that this requires
a detailed knowledge of the charge distribution.\\ \indent 
The density of states (\ref{eq42}) has been obtained assuming
a perturbation which is (in a mathematical sense) asymptotically
far away from the sample. The derivative with respect to the energy $E$
is a consequence of the asymptotic nature of this perturbation.
Physically what counts is the density of states in a finite volume 
$\Omega$. To obtain these densities it is better to first calculate 
the local densities corresponding to Eq. (\ref{eq43}) and to integrate
these local densities over the volume $\Omega$. As shown below the local
densities are not given by energy derivatives of the scattering matrix
but by functional derivatives with respect to the local potential 
$eU({\bf r}).$ The densities determined by integration of such local 
densities of states
differ in general from a simple energy
derivative by a quantum correction \cite{Aronov,Gasp}.
The difference vanishes 
in the semi-classical (WKB) limit. The same is of course
valid for the derivatives in Eq. (\ref{eq32}). \\ \indent
We conclude this section with a remark on localized states.
Equation (\ref{eq42}) is not complete. A conductor might
also contain a contribution to the density of states from
localized states, in addition to
the extended scattering states considered so far.
For the external response the localized states play no role, but
later when we consider the screening the localized states are important.  
To be brief, however, we do not discuss here their role in detail.

\subsection{Response to an oscillating electrostatic potential}
\label{response}
To investigate the self-consistency of dc-transport we already
discussed the local charge distribution inside the conductor.
In Eq. (\ref{eq08}) we introduced a {\em local partial density of
states} which we called the injectivity (see Fig. 6).
Now we introduce additional
local partial densities of states which permit eventually to write
the dynamical conductivity in a simple and transparent manner.
We are interested in the currents generated at the contacts
of a sample in the presence of an oscillating
potential
$\delta U({\bf r}, t).$
We can Fourier transform this potential
with respect to time
and consider a perturbation of the form
\begin{eqnarray}
\delta U({\bf r}, t) = u({\bf r}) (U_{+\omega} \exp(-i\omega t) +
U_{-\omega} \exp(+i\omega t))\;\;. 
\label{eq46}
\end{eqnarray}
Since the potential is real
we have $U_{-\omega}= U^{\ast}_{+\omega}$.
The response to such a potential can be treated using
a scattering approach \cite{MBHTP}: due to the oscillating
internal potential a carrier incident with energy $E$
can gain or loose modulation energy $\hbar \omega$
during reflection at the sample or during transmission
through the sample. The amplitude of an outgoing wave is
a superposition of carriers incident at energy $E$ and
at the side-band energies, $E \pm \hbar \omega.$
In the low-frequency limit the amplitudes of the out going waves
can be obtained by considering the scattering matrix
${\bf s}_{\alpha\beta} (U({\bf r}, t), E)$
to be a slowly varying function of the potential
$U({\bf r}, t).$
Since the deviations of the actual potential away from the
(time-independent) equilibrium potential
$U_{eq}({\bf r})$ are small, we can expand
the scattering matrix in powers of
$\delta U({\bf r}, t) =
U({\bf r}, t) -
U_{eq}({\bf r})$ to linear order
\begin{eqnarray}
{\bf s}_{\alpha\beta}
(U({\bf r}, t), E) =
{\bf s}_{\alpha\beta}
(U_{eq}({\bf r}), E) +
(\delta {\bf s}_{\alpha\beta}/
\delta U({\bf r}))
\delta U({\bf r}, t)\;\;.
\label{eq47}
\end{eqnarray}
\begin{figure}
\vspace{6cm}
\caption{Decomposition of the local density of states of a two 
terminal conductor into injectivities. After Ref. 13.}
\end{figure}
Evaluation of the current at contact $\alpha$ gives \cite{MBHTP}
\begin{eqnarray}
\delta I_{\alpha}^{i}(\omega) =  i e^{2} \omega
\int d^{3}{r} (dn(\alpha, {\bf r})/dE)
u({\bf r})
U_{+\omega}\;\;.
\label{eq48}
\end{eqnarray}
Here we have introduced the local partial density of states \cite{MBCAP,MBHTP}
\begin{eqnarray}
\frac{dn(\alpha ,{\bf r})}{dE} =
- \frac{1}{4 \pi i} \sum_{\beta}
Tr\left[{\bf s}^{\dagger}_{\alpha\beta}
  \frac{\delta {\bf s}_{\alpha\beta}}{e\delta U({\bf r})}
- \frac{\delta {\bf s}^{\dagger}_{\alpha\beta}}{e\delta U({\bf r})}
{\bf s}_{\alpha\beta} \right] 
\label{eq49}
\end{eqnarray}
which we call the {\em emissivity} (see Fig. 7). It describes
the local density of states of carriers at point
${\bf r}$ which are emitted by the conductor at probe $\alpha$. 
A more detailed derivation of Eq. (\ref{eq48}) can be found in
Ref. \cite{MBHTP}. It is useful to express the response to the
internal potential in the from of a conductance defined as
$\delta I_{\alpha}^{i}(\omega) = G^{i}_{\alpha}(\omega)
U_{+\omega}$.
Comparison with
Eq. (\ref{eq45}) gives for the internal conductances
\begin{eqnarray}
G^{i}_{\alpha}(\omega) = i e^{2}
\omega
\int d^{3}{r} \: (dn(\alpha, {\bf r})/dE)\: u({\bf r}) \;\;.
\label{eq50}
\end{eqnarray}
Below we use this internal response to complete the calculation of
the total current. Before doing this it seems useful to pause
for a moment and to discuss in more detail
the local density of states which determine
the internal response.

\subsection{Decomposition of the local density of states}
\label{locpart}
In Eq. (\ref{eq09}) we have expressed the injectivity (\ref{eq08})
with the help of scattering states. Now we give an expression 
of the injectivity in terms of derivatives of the scattering matrix.
Expressions which relate wave functions
to functional derivatives are known from the discussion of the
characteristic times occurring in tunneling processes \cite{LEVE,EAVES}.
Consider for a moment a one-dimensional 
scattering problem with a potential $V(x)$ in an interval $(-a, a)$.
The scattering matrices are $s_{\alpha\beta}$ 
where $\alpha$ and $\beta$ take the values
$1$ and $2$ to designate left and right, respectively. 
Of interest is the time a particle {\it dwells} in this region irrespective
of whether it is ultimately transmitted or whether it is ultimately
reflected. There are two dwell
times $\tau_{D \alpha}$, for the particles
arriving from the left or from the right.
In terms of the scattering states $\Psi_{\alpha}(x)$
and the incident current $I$, the dwell time is given by \cite{EAVES} 
\begin{eqnarray}
\tau_{D \alpha} = \int_{x}^{x+a} dx |\Psi_{\alpha}(x)|^{2}/I \;\;.
\label{eq51}
\end{eqnarray}  
To find the time a particle
dwells in an interval $(x, x+a)$ an infinitesimal uniform
perturbation $dV$ is added in this region to the potential $V(x)$. 
It is found that the dwell time is then related to the 
scattering matrix via the following relationship \cite{LEVE,EAVES}
\begin{eqnarray}
\tau_{D \alpha}= \hbar \: {\rm Im} \left( \: |s_{1\alpha}|^{2} 
\frac{d\ln s_{1\alpha}}{dV} + |s_ {2\alpha}|^{2}
\frac{d\ln s_{2\alpha}}{dV}\: \right) \;\; .
\label{eq52}
\end{eqnarray}
where Im denotes the imaginary part.
For a plane-wave scattering state with wave vector $k$ the current is
$v=\hbar k/m$. Thus a comparison with Eq. (\ref{eq09})
shows that for a single quantum channel the 
dwell time is related to the injectivity by
\begin{eqnarray}
\tau_{D \alpha}/h= \int_{x}^{x+a} dx \: (dn(x,\alpha)/dE)\;\; .
\label{eq53}
\end{eqnarray}
This means that their exists a simple relationship between local 
density of states and derivatives of the scattering matrix
with respect to potentials. It is easy to extend this relation
to the case of an arbitrary space dependent potential and to an 
arbitrary number of channels. The final result is that 
the injectivity is given by \cite{MBCAP,MBHTP}
\begin{eqnarray}
dn({\bf r},\alpha)/dE =
- \frac{1}{4 \pi i} \sum_{\beta}
Tr\left[{\bf s}^{\dagger}_{\beta\alpha}
\frac{\delta {\bf s}_{\beta\alpha}}{e\delta U({\bf r})}
- \frac{\delta {\bf s}^{\dagger}_{\beta\alpha}}{e\delta U({\bf r})}
{\bf s}_{\beta\alpha}\right]\;\;.
\label{eq54}
\end{eqnarray}
\begin{figure}
\vspace{6cm}
\caption{Decomposition of the local density of states of a two 
terminal conductor into emissivites.}
\end{figure}
The injectivity contains information about the origin of the particles:
it is important through which contact the carriers enter
the conductor. Consequently, 
the summation is over the first index of the scattering matrices.
In contrast, the {\em emissivity} defined in Eq. (\ref{eq49})
contains information about the future of the carriers: it is important
through which contact the carriers leave the sample. The summation
is thus over the second index of the scattering matrix.
The sum of all the injectivities (see Fig. 6) or the sum of all the 
emissivities (see Fig. 7)
is equal to the local density of states,
\begin{eqnarray}
dn({\bf r})/dE = \sum_{\alpha} dn(\alpha,{\bf r})/dE = \sum_{\alpha} dn({\bf r},\alpha)/dE
\;\;.
\label{eq55}
\end{eqnarray}
Equation (\ref{eq55}) represents a decomposition of the local
density of states into emissivities and injectivities.
We mention that the emissivities and injectivities are not
independent of one another.
In fact, in the absence of a magnetic field they are equal,
$dn(\alpha,{\bf r})/dE = dn({\bf r},\alpha)/dE$.
In the presence of a magnetic field the microreversibility of the 
scattering matrix implies that the emissivity into contact
$\alpha$ in magnetic field $B$
is equal to the injectivity of contact $\alpha$ if the magnetic field 
is reversed,
\begin{eqnarray}
dn(\alpha ,{\bf r};B)/dE = dn({\bf r},\alpha;-B)/dE\;\; .
\label{eq56}
\end{eqnarray}
While the local density of states at equilibrium is an even function of 
the magnetic field, i. e. $dn( {\bf r};B)/dE = dn({\bf r} ;-B)/dE$, 
the injectivities
and emissivities are in general {\em not} even functions of $B$. This has some
peculiar physical consequences, as has been shown recently by a
low-frequency measurement of capacitances in a quantum Hall system
\cite{CHEN}.

\subsection{Combined external and internal response}
\label{combined} 
We need to find an expression of the current response generated by the 
electric potential oscillations caused by the external potentials 
$\delta \mu_{\alpha} (\omega)$ $ \exp(-i\omega t)$.
We are interested in the response to
first order in $\omega$, and since the currents in Eqs.
(\ref{eq44}) and (\ref{eq48}) are proportional to  $\omega$
it is sufficient to know the quasi-static nonequilibrium state
discussed in Sect. \ref{potential}. We express the
deviation of the potential away from the equilibrium potential 
with the help of the characteristic potentials 
defined in Eq. (\ref{eq06}).
>From Eq. (\ref{eq48}) we find
\begin{eqnarray}
\delta I_{\alpha}^{i}(\omega) =  i e\omega \sum_{\beta} 
\int d^{3}{r} \: (dn(\alpha , {\bf r})/dE) \: 
u_{\beta}({\bf r})\delta \mu_{\beta }(\omega)\;\; .
\label{eq57}
\end{eqnarray}
Thus the induced potentials give rise to a conductance  
\begin{eqnarray}
G^{i}_{\alpha \beta} (\omega) = i e^{2} \omega 
\int d^{3}{r}\: (dn(\alpha , {\bf r})/dE) \: u_{\beta}({\bf r})
\:\:,
\label{eq58}
\end{eqnarray}
which can be written in terms the Green's function and the
injectivity,
\begin{eqnarray}
G^{i}_{\alpha \beta} (\omega) = 
i e^{2} \omega 
\int d^{3}{ r} 
\int d^{3}r^{\prime}
(dn(\alpha, {\bf r})/dE)
g ({\bf r}, {\bf r}^{\prime})
(dn({\bf r}^{\prime},\beta)/dE)\;\;.
\label{eq59}
\end{eqnarray}
Eq. (\ref{eq59}) tells us that the internal response 
$G^{i}_{\alpha \beta} (\omega)$ is a 
consequence of the charge injected from contact 
$\beta$ which generates a potential 
determined by the Green's function and that this potential in turn
generates a current at $\alpha$ determined by the emissivity
into that contact.
The total response is the sum of the external response (\ref{eq45})
and the internal response (\ref{eq59}),
\begin{eqnarray}
G_{\alpha \beta} (\omega) = G^{e}_{\alpha \beta} (\omega)  
+ G^{i}_{\alpha \beta} (\omega)\;\;.
\label{eq60}
\end{eqnarray}
We express it in the form 
\begin{eqnarray}
G_{\alpha \beta} (\omega) = G_{\alpha \beta} (0)  
-i \omega E_{\alpha \beta}  + O(\omega^{2})
\label{eq61}
\end{eqnarray} 
and call $E_{\alpha \beta}$ the (screened) {\it emittance} of the
conductor. It is given by \cite{MBCAP}
\begin{eqnarray}
E_{\alpha \beta} = e^{2}\frac{dN_{\alpha\beta}}{dE} -
e^{2} \int d^{3}{ r} \int d^{3}r^{\prime}\:
\frac{dn(\alpha, {\bf r})}{dE}\:
g ({\bf r}, {\bf r}^{\prime}) \:
\frac{dn({\bf r}^{\prime},\beta)}{dE}.
\label{eq62}
\end{eqnarray}
Before we apply this result in the next section to a few simple
problems, we want to demonstrate that $G_{\alpha \beta}$
is indeed current conserving, i. e.
that the rows and columns of the emittance matrix add up to zero. 
Consider the first column. If we add $E_{11}$ and $E_{21}$ the 
first terms in the emittance give the total charge 
$dN_{11}/{dE} + dN_{21}/{dE}$ injected from contact $1.$
In the second term the two emissivities add to give
the local density of states. Now Eq. (\ref{eq16})
is used. What remains is the integral over the entire volume 
of the injectivity which is just the total injected charge. 
Thus for a two terminal conductor in electrical isolation
the emittance matrix satisfies
$E_{\mu} \equiv E_{11} = -E_{12} = -E_{21} = E_{22}$.

\section{Examples}
\label{examples}
\subsection{Emittance of a metallic diffusive conductor}
\label{metallic}
Let us consider a mesoscopic metallic conductor connecting
two reservoirs. In a metallic conductor charge accumulations
are screened over a Thomas-Fermi screening length (apart from
miniscule and more subtle Friedel-like long-range effects \cite{LE89}).
If we assume in addition that the density varies not to rapidly 
then the local potential is directly determined by the local density.
The local potential $\delta U({\bf r})$ generated by an injected 
charge $\delta n_{in}({\bf r})$ is determined by 
$(dn({\bf r})/dE) \: e\delta U({\bf r}) = \delta n_{in}({\bf r}).$
This corresponds to a Green's function which is  
a delta function in space and with a weight 
inversely proportional to the local density of states 
$ g({\bf r}, {\bf r}^{\prime}) = (dn({\bf r})/dE)^{-1}
\delta ({\bf r}-{\bf r}^{\prime})$.
Using this in Eq. (\ref{eq62}) gives an emittance in terms of 
densities only \cite{MBHTP},
\begin{equation}
E_{\alpha \beta} (\omega) = e^{2}\frac{dN_{\alpha\beta}}{dE} -
e^{2} \int d^{3}{r} \: 
\frac{dn(\alpha, {\bf r})}{dE}\:
(\frac{dn({\bf r})}{dE})^{-1} \:
\frac{dn({\bf r},\beta)}{dE} \;\;.
\label{eq63}
\end{equation}
There are no electric field lines outside the conductor.\\ \indent
The wire with cross-section $A$ ranges from $x=-L/2$ to
$x=L/2$. The mean distance between the impurities is $l$.
The reflection and the transmission probability per channel are  
$R=1-l/L$ and $T=l/L$, respectively. The partial densities of
states of reflected carriers are 
$dN_{11}/dE = dN_{22}/dE = (1/2)  (1-l/L) (dN/dE)$ where
$dN/dE$ is the total density of states in the volume 
$\Omega = AL$. The partial densities of states of transmitted carriers
are $dN_{12}/dE = dN_{21}/dE = (1/2)
(l/L) (dN/dE)$. The diffusion equation for the diffusive metallic
conductor implies
the ensemble averaged and over the cross section averaged
injectivities $dn(x,1)/dE = (1/2L)(dN/dE) (1- 2x/L)$ and 
$dn(x,2)/dE = $ $(1/2L)$ $(dN/dE)$ $(1+ 2x/L)$.
In the absence of a magnetic field, the emissivities are given by the
same expressions. The linear dependence of the injectivities gives
an ensemble averaged potential which drops also linearly.
With the help of these expressions, we find for the emittance \cite{MBUNP}
\begin{equation}
E_{\mu} = (1/6 -l/2L) (dN/dE) \;\;.
\label{emmet}
\end{equation}
Already for metallic wires longer then
$3l$ the emittance is a positive quantity.
Since typically $l/L$ is much smaller than $1/3$ the metallic
diffusive wire at low frequencies responds
like a capacitor with an ensemble averaged capacitance
given by Eq. (\ref{eqI2}). 

\subsection{Emittance of a perfect ballistic wire}
\label{ballistic}
Consider a ballistic wire of length L with N quantum channels.
We apply again Thomas-Fermi screening. However, this is not
very well justified and permits to obtain an estimate only.
We also ignore the variation of the potentials
near the contacts, which in a more realistic treatment might well give
us a capacitive contribution. Each quantum channel contributes 
with a density of states $2L/hv_{n}$, were 
$v_{n}$ is the channel velocity evaluated at the Fermi energy.
The total density of states per spin is
$dN/dE = \sum_{n} 2L/hv_{n}$. In the absence of backscattering
it holds 
$dN_{11}/dE = dN_{22}/dE = 0$ and $dN_{21}/dE =dN_{12}/dE = (1/2)\:
dN/dE$. The injectivities $dn(x,1)/dE = (1/2L)(dN/dE)$ and
$dn(x,2)/dE = (1/2L)(dN/dE)$ are independent of the space coordinate.
This corresponds to a potential in the ballistic wire which is constant
and midway between the electrochemical potentials at the contacts.
Hence, using Eqs. (\ref{eq62}) we find that an ideal perfect wire
has a negative emittance given by Eq. (\ref{eqI3}).
A ballistic wire responds like a 
conductor which classically is represented by a resistance and 
an inductance in series, and where the emittance can be viewed as a 
kinetic inductance.

\subsection{Emittance of a resonant double barrier}
\label{barrier}
As a third example we consider a resonant double barrier \cite{MBHTP}.
The scattering matrix is 
$s_{\alpha\beta} = (\delta_{\alpha\beta}-i \Gamma_{\alpha} \Gamma_{\beta}
/\Delta)\exp(i\delta_{\alpha}+ i\delta_{\beta})$, where
$\delta_{\alpha\beta}$
is the Kronecker symbol and the $\delta_{\alpha}$ are phases whose energy
dependence can be neglected compared to the rapid variation of the
resonant denominator $\Delta = E - E_{r} - e\delta U - i \Gamma /2$
with $\Gamma = \Gamma_{1}+\Gamma_{2}$.
Here, $E_{r}$ is the resonant energy at equilibrium and $e\delta U$ is
the deviation of the electrostatic potential 
away from its equilibrium value at the site of the long lived state in
the presence of transport. 
With the help of the injectances $dN_{\alpha}/dE = 
(\Gamma _{\alpha}/2 \pi |\Delta|^{2})$ and the total density of states
$dN/dE = 
(\Gamma/2 \pi |\Delta|^{2})$
the partial densities of 
state can be expressed in the following manner:
For $\alpha = \beta $ we find 
\begin{equation}
\frac{dN_{\alpha\alpha}}{dE} =  \frac{R}{2} \frac{dN}{dE}
\pm \frac{1}{2}(\frac{dN_{1}}{dE} -\frac{dN_{2}}{dE}) 
\label{eq66}
\end{equation}
where the plus and the minus sign correspond to $\alpha = 1$
and $\alpha = 2$, respectively.
Here, $R=1-T$ is  
the reflection probability. 
For $\alpha \neq \beta$,
one has
\begin{equation}
\frac{dN_{\alpha \beta}}{dE} =  \frac{T}{2}\frac{dN}{dE}\;\;. 
\label{eq67}
\end{equation} 
The unscreened injectances (emittances)
are found by integrating the injectivities (emissivities)
over the volume of the localized state, i. e. over the well.
They are given by  
\begin{eqnarray}
dN_{1\alpha}/dE + dN_{2\alpha}/dE & = & dN_{\alpha}/dE\;\; .
\label{eq70}
\end{eqnarray}
In a Thomas-Fermi approach the characteristic potentials 
$u_{\alpha}$ in the well are 
determined by $(dN/dE) u_{\alpha} = dN_{\alpha}/dE .$
This gives $u_{\alpha} = \Gamma_{\alpha}/\Gamma .$
Using Eq. (\ref{eq63}) gives an emittance \cite{MBHTP}
\begin{equation}
E_{\mu} = - e^{2} \: \frac{(dN_{1}/dE)\:(dN_{2}/dE)}{dN/dE}
\left( 
\frac{\Gamma^{2} /2 - |\Delta|^{2}}{|\Delta|^{2}}\right)
\label{eq71}
\end{equation}
For a symmetric resonant tunneling barrier Eq. (\ref{eq71}) simplifies
and is given by Eq. (\ref{eqI4}). 
At resonance the emittance is negative reflecting kinetic
(inductive) behavior, it is zero at half-width of the resonance, and it is
positive (capacitive) if the Fermi level is more than
a half width above (below) the resonant energy.  
Clearly, Thomas-Fermi screening is not very realistic for such a
conductor. Moreover, the quantization of charge in the well might
play a decisive role. Nevertheless, these considerations indicate the
character of the results that a more realistic treatment might yield and
hopefully stimulate work in that direction.

\section{Summary}
\label{summary}
We have developed a self-consistent discussion of mesoscopic
electrical conduction. The determination of the electrical
potential in the presence of a dc-current, although unimportant for
the discussion of the dc-conductances itself, permits to 
discuss the conditions under which the dc-conductance formulae are
valid, it permits to calculate the first nonlinear corrections
of the purely linear response, and permits to find the ac-conductances
to first order in frequency. We applied the results to some simple
examples.\\ \indent
We have emphasized the case of two terminal conductors in electric
isolation. The theory 
permits, however, also to discuss the effect of nearby capacitors and gates
and in fact provides a mesoscopic description of 
capacitances \cite{MBCAP,MBROM}.
Some implications of a mesoscopic theory of capacitance, like 
Aharonov-Bohm oscillations in capacitance coefficients, 
the gate voltage dependence of persistent currents 
have already been the subject
of recent works \cite{MBREK,MBCS}.\\ \indent
The self-consistent nature of electrical transport is a consequence
of the long-range Coulomb interaction of carriers. A self-consistent
description must, therefore, tackle an interacting many-particle problem.
Consequently, the results of such a theory depend somewhat on the 
sophistication that is used to treat the many-particle problem. Here
we have used a simple Hartree approach. Since density-functional
theory is nothing but an improved Hartree theory it gives results
which look formally very similar to the results presented
here \cite{MBTRI}. A stronger modification of the results discussed here can be
expected in situations where one must take the quantization of charge into 
account \cite{MBCS,BS}.
 
The theory presented here, demonstrates that interesting
results can be obtained by investigating nonlinearities 
and ac-conductances. The theory demonstrates that it is 
necessary to treat nonlinearities and the ac-response 
self-consistently to conserve both charge and 
current. We are confident that experiments will eventually 
demonstrate the close connection between electrostatic questions
and nonlinearities and ac-response.

\end{document}